\documentclass[aps,prl,twocolumn,showpacs,groupedaddress]{revtex4}  
\usepackage{graphicx}  
\usepackage{dcolumn}   
\usepackage{bm}        
\usepackage{amssymb}   
\usepackage{amsmath}

\begin{document}
\hyphenation{EVTGEN} 

\hspace{5.2in} \mbox{Fermilab-Pub-08/518-E}\\
\title{Evidence for the decay $B^0_s\rightarrow D_s^{(*)}D_s^{(*)}$ and a measurement of $\Delta\Gamma_s^{\rm CP}/\Gamma_s$}
%
\author{V.M.~Abazov$^{36}$}
\author{B.~Abbott$^{75}$}
\author{M.~Abolins$^{65}$}
\author{B.S.~Acharya$^{29}$}
\author{M.~Adams$^{51}$}
\author{T.~Adams$^{49}$}
\author{E.~Aguilo$^{6}$}
\author{M.~Ahsan$^{59}$}
\author{G.D.~Alexeev$^{36}$}
\author{G.~Alkhazov$^{40}$}
% ...

\affiliation{\vspace{0.1 in}(The D\O\ Collaboration)\vspace{0.1 in}}
% ...
\affiliation{$^{6}$University of Alberta, Edmonton, Alberta, Canada,
                Simon Fraser University, Burnaby, British Columbia, Canada,
                York University, Toronto, Ontario, Canada, and
                McGill University, Montreal, Quebec, Canada}
% ...
\affiliation{$^{29}$Tata Institute of Fundamental Research, Mumbai, India}
% ...
\affiliation{$^{36}$Joint Institute for Nuclear Research, Dubna, Russia}
% ...
\affiliation{$^{40}$Petersburg Nuclear Physics Institute,
                St. Petersburg, Russia}
% ...
\affiliation{$^{49}$Florida State University, Tallahassee, Florida 32306, USA}
% ...
\affiliation{$^{51}$University of Illinois at Chicago,
                Chicago, Illinois 60607, USA}
% ...
\affiliation{$^{59}$Kansas State University, Manhattan, Kansas 66506, USA}
% ...
\affiliation{$^{65}$Michigan State University,
                East Lansing, Michigan 48824, USA}
% ...
\affiliation{$^{75}$University of Oklahoma, Norman, Oklahoma 73019, USA}
% ...
\date{February 04, 2008}

\begin{abstract}
We search for the semi-inclusive process $B^0_s\rightarrow D_s^{(*)}D_s^{(*)}$ using 2.8 fb$^{-1}$ of $p \bar{p}$ collisions 
at $\sqrt{s} = 1.96$~TeV recorded by the D0 detector operating at the Fermilab Tevatron Collider. 
We observe $26.6\pm8.4$ signal events with a significance above background of 3.2 standard deviations
yielding a branching ratio of ${\cal B}(B^0_s\rightarrow D_s^{(*)}D_s^{(*)}) = 0.035\pm0.010\text{(stat)}\pm0.011\text{(syst)}$. 
Under certain theoretical assumptions, these double-charm final states saturate CP-even eigenstates in the $B_s^0$ decays
resulting in a width difference of $\Delta \Gamma_s^{\rm CP}/\Gamma_s = 0.072\pm0.021\text{(stat)}\pm0.022\text{(syst)}$. 
\end{abstract}

\maketitle
\newpage

The phenomenon of CP violation is believed to be intimately tied to explaining 
the matter dominance in the present day universe~\cite{sakarov}. 
CP violation is expected to occur in the evolution of neutral particles that can mix between different eigenbases. 
For the $B_s^0$ system, the flavor eigenstates can be decomposed into 
heavy ($H$) and light ($L$) states based on mass or into even and odd states based on CP. 
The width differences between these eigenstates are defined by $\Delta\Gamma_s = \Gamma_L - \Gamma_H$ 
and $\Delta\Gamma_s^{\rm CP} = \Gamma_s^{\rm even} - \Gamma_s^{\rm odd}$, respectively. 
These two quantities are connected with the possible presence of new physics (NP) by $\Delta\Gamma_s = \Delta\Gamma_s^{\rm CP}\cos\phi_s$, 
where $\phi_s$ is the CP violating mixing phase which constrains models of NP.

In the standard model (SM) a mixing parameter, $\Gamma_{12}$, determining the size of the width difference between CP eigenstates
stems from the decays into final states common to both $B$ and $\bar{B}$. 
Since this quantity is dominated by CKM-favored tree-level decays, it is practically insensitive to NP. 
Due to the hierarchy of the quark mixing matrix~\cite{wolfenstein}, 
the width difference is governed by the partial widths of $B_s^0$ decays into final CP eigenstates 
through the $b\rightarrow c\bar{c}s$ quark-level transition, such as $B^0_s\rightarrow D_s^+D_s^-$ or $B^0_s\rightarrow J/\psi\phi$.
Topologically, the former type of decay mode is a color-allowed spectator, 
while the latter type is suppressed by the effective color factor.
Thus, the semi-inclusive decay modes $B^0_s\rightarrow D_s^{(*)}D_s^{(*)}$, 
where $D_s^{(*)}$ denotes either $D_s^{\pm}$ or $D_s^{\pm *}$, are interesting 
because they give the largest contribution to the difference between the widths of the heavy and light states.
The other decay modes are estimated to contribute less than 0.01 to the projected $\sim 0.15$ value of $\Delta\Gamma_s/\Gamma_s$~\cite{assumption},
where $\Gamma_s(=1/\tau_s)\equiv (\Gamma_L+\Gamma_H)/2$.

In the Shifman-Voloshin (SV) limit~\cite{shifman_voloshin}, 
given by $m_b-2m_c \rightarrow 0$ with $N_c\rightarrow\infty$ (where $N_c$ is the number of colors), 
$\Delta \Gamma_s^{\rm CP}$ is saturated by $\Gamma(B^0_s\rightarrow D_s^{(*)}D_s^{(*)})$.
Then the width difference can be related to the branching ratio of $B_s^0$ mesons 
to this inclusive double-charm final state by \cite{B@Tev, pursuit}
\begin{multline}
2{\cal B}(B_s\rightarrow D_s^{(*)}D_s^{(*)})\\
 \simeq \Delta\Gamma_s^{\rm CP}\left[\frac{\frac{1}{1-2x_f}+\cos\phi_s}{2\Gamma_L}+\frac{\frac{1}{1-2x_f}-\cos\phi_s}{2\Gamma_H}\right],
\label{br_delgam}
\end{multline}
where $x_f$ is the fraction of the CP-odd component of the decay, $\Gamma_s^{\rm odd}/\Gamma_s^{\rm even}=x_f/(1-x_f)$. 
Therefore, given the CP structure of the final state, 
$\Delta\Gamma_s^{\rm CP}$ can be measured using the information from branching ratios without lifetime fits. 
The irreducible theoretical uncertainty of this approach stems from the omission of CKM-suppressed decays 
through the $b\rightarrow u\bar{u}s$ transition which is of order $2|V_{ub}V_{us}/V_{cb}V_{cs}|\sim3-5\%$. 

In this Letter we report the first evidence for the decay $B_s^0\rightarrow D_s^{(*)}D_s^{(*)}$.
The study uses a data sample of $p\bar{p}$ collisions at $\sqrt{s}=1.96$ TeV corresponding to an integrated 
luminosity of $2.8$ fb$^{-1}$ recorded by the D0 detector operating at the Fermilab Tevatron Collider during 2002 - 2007. 
This supersedes our previous study of the same final state based on $1.3$ fb$^{-1}$~\cite{james}. 
A similar study based on events containing two $\phi$ mesons has been reported 
by the ALEPH collaboration at the CERN LEP Collider~\cite{ALEPH}.

This analysis considers the $B_{s}^0$ decay into two $D_s^{(*)}$ mesons. 
No attempt is made to identify the photon or $\pi^0$ emanating from the $D_s^*$ decay.
We search for one hadronic $D_s$ decay to $\phi\pi$ and one semileptonic $D_s$ decay to $\phi\mu\nu$, 
where both $\phi$ mesons decay to $K^+K^-$. 
The branching fraction is extracted by normalizing the $B_s^0\rightarrow D_s^{(*)}D_s^{(*)}$ decay 
to the $B_{s}^0 \rightarrow D_s^{(*)}\mu\nu$ decay.

D0 is a general purpose detector \cite{D0_detector} consisting of a central tracking system, 
uranium/liquid-argon calorimeters, and an iron toroid muon spectrometer. 
The central tracking system allows charged particles to be reconstructed. 
This system is composed of a silicon microstrip tracker (SMT) and a central fiber tracker (CFT) 
embedded in a 2 T solenoidal magnetic field.
Muons are identified and reconstructed with a magnetic spectrometer located outside of the calorimeter. 
The spectrometer contains magnetized iron toroids and 
three super-layers of proportional drift tubes along with scintillation trigger counters. 
Information from the muon and tracking systems is used to form muon triggers.
For the events used by this analysis, 
the muon from the semileptonic $D_s$ decay satisfies the inclusive single-muon triggers.

Muons are identified by requiring segments reconstructed in at least two out of the three super-layers in the muon system 
and associated with a trajectory reconstructed with hits in both the SMT and the CFT. 
We select muon candidates with transverse momentum $p_T > 2.0$ GeV/c and total momentum $p > 3.0$ GeV/c.

$\phi$ mesons are formed from two opposite sign charged particles with $p_T > 0.7$ GeV/c in the event assuming a kaon mass hypothesis.
We require at least one kaon to have an impact parameter clearly separated from the $p\bar{p}$ interaction point (primary vertex) 
with at a minimum 4 standard deviations significance. 
The two-kaon systems satisfying $p_T(KK) > 2.0$ GeV/c and $1.010<m(KK)<1.030$ GeV/c$^2$ are selected as $\phi$ candidates.

The hadronic $D_s$ meson is reconstructed by combining the $\phi$ candidate 
with a third track with $p_T > 0.5$ GeV/c which is assigned the pion mass.
The pion is required to have charge opposite to that of the muon.
The three particles must form a well reconstructed vertex displaced from the primary vertex~\cite{vtx_algo}.
We require the cosine of the angle between the $D_s$ momentum 
and the direction from the primary vertex to the $D_s$ vertex to be greater than 0.9.
For the signal decay chain of a pseudoscalar to a vector plus pseudoscalar, followed by the decay of the vector to two pseudoscalars, 
$\cos\theta_{\phi}$ is distributed quadratically, 
where $\theta_{\phi}$ is the decay angle of a kaon in the $\phi$ rest frame with respect to the direction of the $D_s$ meson, 
and hence a constraint $|\cos\theta_{\phi}|>0.3$ is imposed. 

\begin{figure*}[t]
\centering
\begin{minipage}[t]{0.33\textwidth}
\includegraphics[width=0.94\textwidth]{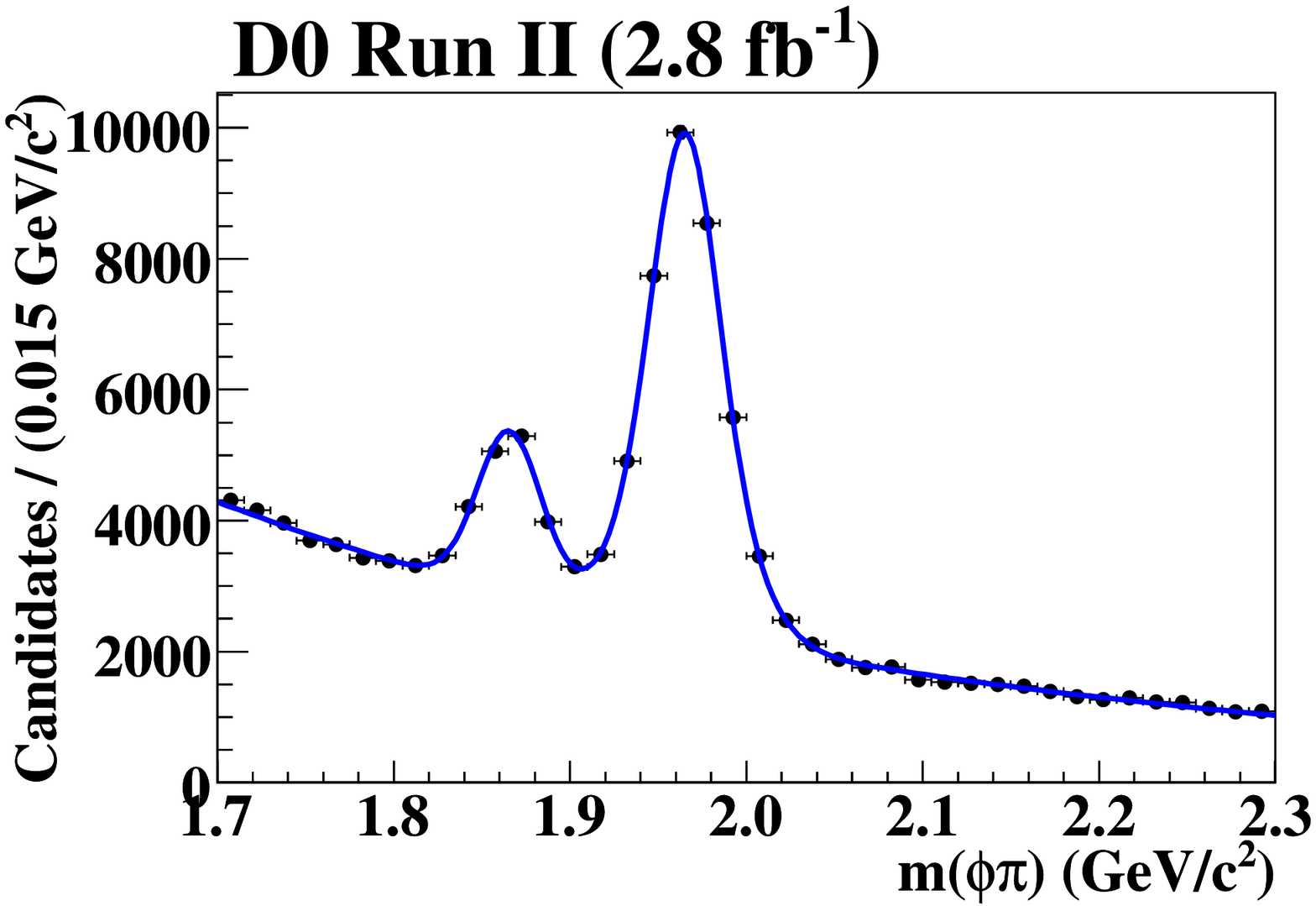}
\caption{Invariant mass distribution of the $\phi\pi$ system for the $B_s^0 \rightarrow D_s^{(*)}\mu\nu$ sample. 
The two peaks correspond to the $D^{\pm}$ candidates (lower masses) and $D_s$ candidates (higher masses).}
\label{mass_dsmu}
\end{minipage}
\hfill
\begin{minipage}[t]{0.63\textwidth}
\hfill
\includegraphics[width=0.495\textwidth]{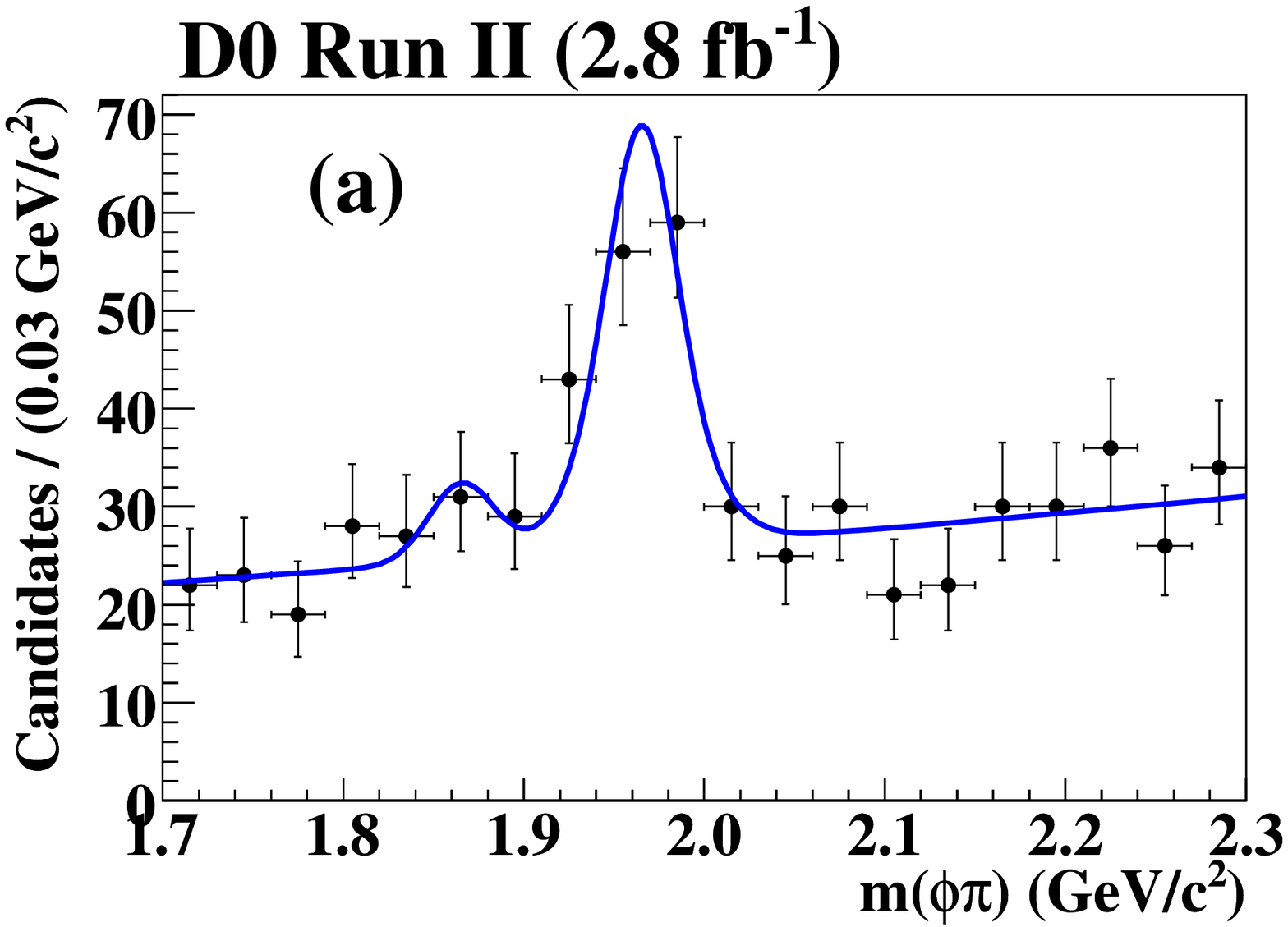}
\includegraphics[width=0.495\textwidth]{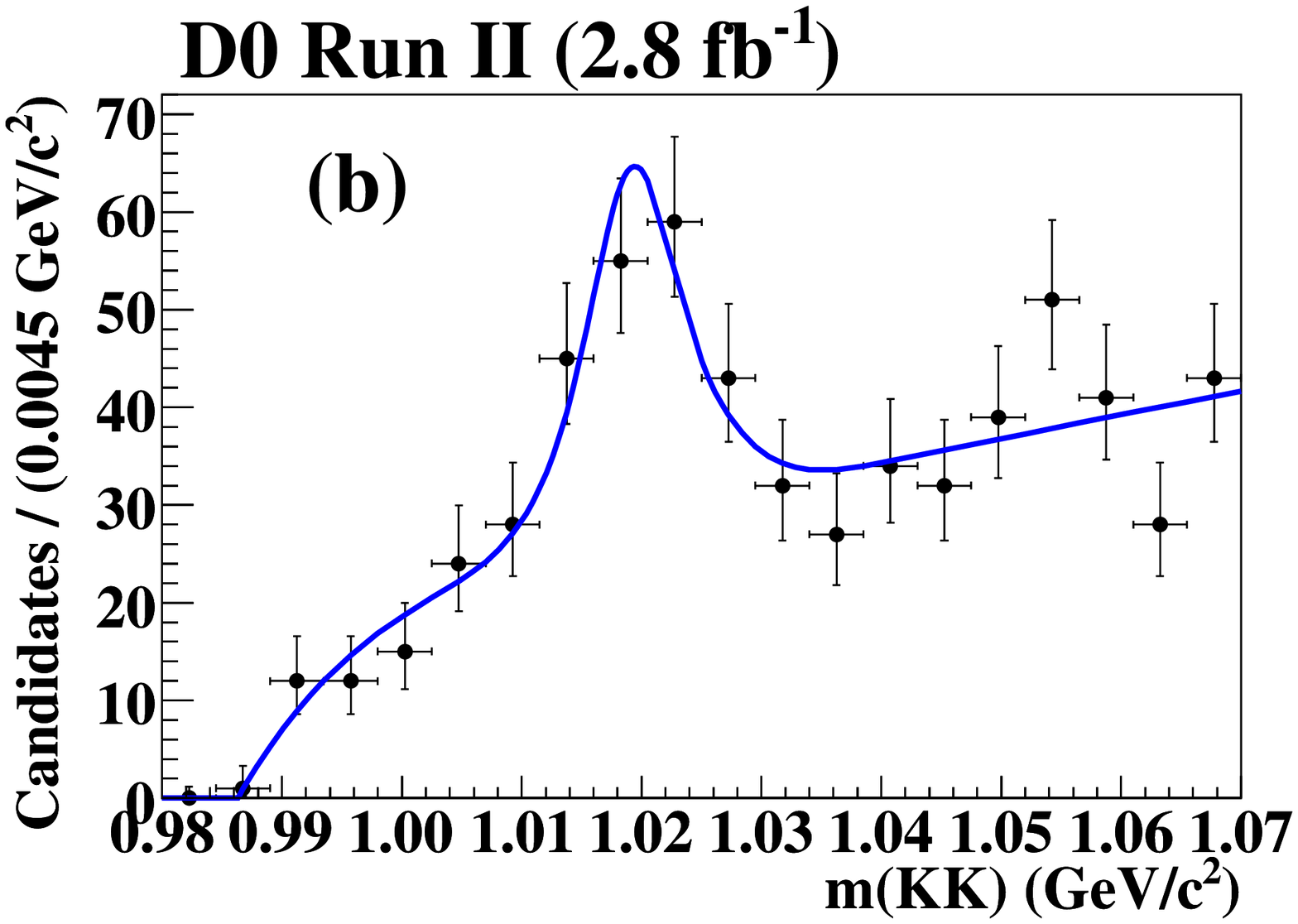}
\caption{Projections of the two-dimensional maximum likelihood fit onto invariant mass spectra of the 
(a) $\phi\pi$ system from hadronic $D_s$ decays and (b) $KK$ system from semileptonic $D_s$ decays. 
The peaks in both distributions are explored to search for the correlation between the two systems.}
\label{mass_dsds}
\end{minipage}
\end{figure*}

The $B_s^0 \rightarrow D_s^{(*)}\mu\nu$ decay vertex is reconstructed based on 
the momentum and direction of the reconstructed hadronic $D_s$ candidate and its intersection with the track of an oppositely charged muon. 
This vertex is required to be located between the primary vertex and the $D_s$ vertex, 
whereby the individual $B_s$ and $D_s$ vertex displacements are consistent with a $p\bar{p}\rightarrow B_s\rightarrow D_s$ decay chain.
The invariant mass of the $B_s^0$ candidate is required to be less than 5.2 GeV/c$^2$.
We require the daughter particles of the $B_s^0$ meson to be well isolated from other tracks.
Background is further suppressed using a likelihood ratio technique~\cite{ctag} that combines information from 
the invariant masses and momenta of the reconstructed particles, vertex quality, and the $\phi$ helicity angle.

The $\phi\pi$ invariant mass distribution for $B_s^0\rightarrow D_s^{(*)}\mu\nu$ candidates is shown in Fig.~\ref{mass_dsmu}. 
Maxima corresponding to the $D_s\rightarrow \phi\pi$ decay and the $D^{\pm}\rightarrow \phi\pi$ decay are clearly observed. 
The $D_s$ signal originates from $\sim 90\%$ semileptonic $B_s^0$ decays 
and $\sim10\%$ decays of the type $B\rightarrow D_sD$ followed by semileptonic $D$ decay.
These fractions are determined from Monte Carlo (MC) simulation 
using the known or estimated branching fractions from the PDG~\cite{PDG} or {\sc evtgen}~\cite{EvtGen}. 
Approximately $2\%$ of the events are due to direct charm production $p\bar{p}\rightarrow DD$, 
determined by using full simulation and reconstruction of $DD^*$ candidates.
The overall sample composition is verified using studies of the
$B$ lifetime and mixing parameters~\cite{lifetime,mixing}.

For the second $\phi$ candidate, we search for an additional pair of oppositely charged particles 
in the event imposing the same criteria as for the first $\phi$ meson.
The two kaon tracks are combined with the muon track to produce a common vertex for the semileptonic $D_s$ candidate. 
We require the $D_s$ candidate to originate from a common vertex to the hadronic $D_s$ candidate 
to complete the $B_s^0\rightarrow D_s^{(*)}D_s^{(*)}$ decay. 
This approach is justified since the average transverse decay length of the $D_s$ meson 
relative to the $B_s^0$ meson decay vertex is $\sim1.0$ mm with an uncertainty of $\sim0.6$ mm. 
By applying the same selection criteria as in the normalization $B_s^0 \rightarrow D_s^{(*)}\mu\nu$ decay sample, 
many detector related systematic effects cancel. 
The total invariant mass is required to lie between 4.30 and 5.20 GeV/c$^2$.

Correlated production of this double-charm decay, 
where both $D_s$ mesons originate from the same parent $B_s^0$ meson, 
is then determined by examining the two-dimensional distribution of $m(\phi\pi)$ 
from hadronic $D_s$ candidates versus $m(KK)$ from semileptonic candidates.
We perform a maximum likelihood fit to this distribution with four components: 
the correlated $D_sD_s$ component is modeled as the product of signal terms in both dimensions, 
the uncorrelated components are modeled as the product of the signal term in one dimension 
and the background term in the other dimension, 
and the background correlation is modeled as the combination of the background terms in both dimensions. 
Signal and background models are expected to be identical with those for the $B_s^0\rightarrow D_s^{(*)}\mu\nu$ sample, 
from which the parameters of the signal models are determined.
Projections of the two-dimensional likelihood fit onto both axes are displayed in Fig.~\ref{mass_dsds}.
The fit returns a yield of $31.0\pm 9.4$ correlated events.

Three possible sources of background are considered in the correlated sample.
Direct charm production from $p\bar{p}$ is estimated based on the fraction of prompt charm 
measured directly in the inclusive $D_s^{(*)}\mu\nu$ sample, $(10.3 \pm 2.5)\%$,
along with the decay fraction of the second charm quark to a $D_s$ meson and the reconstruction efficiency for this decay. 
Due to a shorter decay length of the charm decay, 
the lifetime requirement reduces its contribution significantly leading to an estimate of $(1.9 \pm 0.5)\%$.

The second background source arises from the semileptonic $B_s^0\rightarrow D_s^{(*)}\phi\mu\nu$ decay. 
This can be extracted by studying the $m(\phi\mu)$ spectrum. 
In this variable, $B_s^0\rightarrow D_s^{(*)}D_s^{(*)}$ events tend towards lower values, 
while $B_s^0\rightarrow D_s^{(*)}\phi\mu\nu$ events tend towards higher values.

The third source consists of $B^{\pm,0}\rightarrow D_s^{(*)}D_s^{(*)}KX$ events.
This background can be extracted by studying the visible mass of all reconstructed daughter particles, $m(D_s\phi\mu)$.
The mass tends to have higher values for $B_s^0\rightarrow D_s^{(*)}D_s^{(*)}$ than for  $B^{\pm,0}\rightarrow D_s^{(*)}D_s^{(*)}KX$.

These backgrounds are estimated with MC samples by repeating the fit in three separate regions 
chosen so that mainly one source contributes to each region in the $m(\phi\mu)-m(D_s\phi\mu)$ plane. 
The separate components, the signal and the two latter backgrounds, 
are then extracted based on the expected distribution over the three regions of the three components. 
We find a signal yield of $26.6 \pm 8.4$ events originating from the $B_s^0\rightarrow D_s^{(*)}D_s^{(*)}$ process 
after subtracting the correlated background events. 

The signal is normalized to the total $B_s^0\rightarrow D_s^{(*)}\mu\nu$ yield 
taking into account the composition of the sample as discussed earlier. 
The reconstruction efficiency ratio between the two samples is estimated from MC to be $0.082\pm 0.015$. 
This small value results from the softer muon momentum spectrum in charm decays as compared to bottom decays. 
The systematic uncertainty in the ratio contains uncertainties from the modeling of the $B_s^0$ momentum spectrum, 
the decay form factors and sample composition, and the trigger and reconstruction efficiencies. 
Our efficiency model is verified by comparing the expected and measured $D_s$ yield 
and the relative $B_s^0\rightarrow D_s^{(*)}D_s^{(*)}$ to $B_s^0\rightarrow D_s^{(*)}\mu\nu$ yields as a function of muon $p_T$.

Using all the above inputs, the branching ratio is measured as
\begin{align*}
{\cal B}&(B_s^0\rightarrow D_s^{(*)}D_s^{(*)})\\
&= 0.035\pm0.010({\rm stat})\pm0.008({\rm exp~syst})\pm0.007({\rm ext}),
\end{align*}
where the ``ext'' uncertainty arises from the external input branching ratios taken from the PDG~\cite{PDG}. 
This uncertainty contributes $\sim45\%$ to the total systematic uncertainty (exp syst{\small $\bigoplus$}ext), 
which leaves room for further improvements in the result. 
The experimental systematic uncertainty accounts for the rest of the total systematic uncertainty, 
containing a $37\%$ component from the reconstruction efficiency ratio, 
$11\%$ from the background estimation, and $4\%$ from the fitting procedure. 
All other uncertainties are $\leq1\%$.

The probability that the total background would fluctuate to the measured event yield or higher 
is evaluated to be $1.2\times10^{-3}$ through pseudo-experiments including systematic uncertainties.
This corresponds to a significance of $3.2$ standard deviations.

\begin{figure}[t]
\centering
\includegraphics[width=0.44\textwidth]{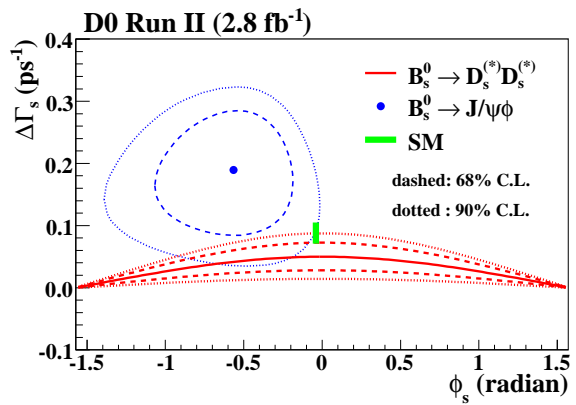}
\caption{Constraints in the $\Delta\Gamma_s-\phi_s$ plane. 
The solid line represents our measurement under the theoretical assumptions stated in the text and with $x_f=0$. 
Two pairs of lines are 68$\%$ (dashed) and 90$\%$ (dotted) C.L. intervals of $\Delta\Gamma_s$ for a given assumed value of $\phi_s$. 
Contours from the $B_s^0\rightarrow J/\psi\phi$ decay are the equivalent C.L. regions 
of ($\Delta\Gamma_s$, $\phi_s$) when measuring simultaneously both parameters. 
No theoretical uncertainties are reflected in the plot. 
The SM prediction is represented by the thick vertical line.}
\label{delgam_phi}
\end{figure}

Information on the mixing-induced CP asymmetry in the $B_s^0$ system can be extracted 
from the branching fraction measurement through Eq.~(\ref{br_delgam}). 
Since the CP structure of the decay is presently not accessible either in theory or experiment, 
several scenarios for different $x_f$ values can be considered. 
In the heavy quark hypothesis~\cite{assumption} along with the SV limit, 
the CP-odd component of the decay vanishes, 
leaving the inclusive final state to be CP-even, i.e. $x_f=0$, 
with a theoretical uncertainty of $\sim5\%$~\cite{HFAG}. 
This scenario is illustrated in Fig.~\ref{delgam_phi}, 
presenting the constraint in the $\Delta\Gamma_s-\phi_s$ plane from this measurement 
assuming the relation $\Delta\Gamma_s=\Delta\Gamma_s^{\rm CP}\cos{\phi_s}$. 
Confidence-level (C.L.) contours from the flavor-tagged decay $B_s^0\rightarrow J/\psi\phi$ at D0~\cite{jpsiphi} are superimposed. 
We take the mean lifetime of $B_s^0$ meson from Ref.~\cite{PDG}.

Furthermore, within the SM framework, the mass eigenstates coincide with the CP eigenstates 
and the expression used in the previous studies~\cite{ALEPH,james} is recovered. 
Our measurement gives
\begin{equation*}
\begin{aligned}
\frac{\Delta\Gamma_s^{\rm CP}}{\Gamma_s} &\simeq \frac{2{\cal B}(B_s^0 \rightarrow D_s^{(*)}D_s^{(*)})}{1-{\cal B}(B_s^0 \rightarrow D_s^{(*)}D_s^{(*)})}\\
\label{delgam_br}
&= 0.072\pm0.021({\rm stat})\pm0.022({\rm syst}).
\end{aligned}
\end{equation*}
This result is consistent with the SM prediction~\cite{theory} as well as with the current world average value~\cite{HFAG}.
Therefore, if the CP structure of the final state can be disentangled and the theoretical errors can be controlled, 
this approach can provide a powerful constraint on mixing and CP violation in the $B_s^0$ system.

In summary, we performed a study of $B_s^0$ decays into the semi-inclusive double-charm final state 
using an integrated luminosity of 2.8 fb$^{-1}$ at the D0 experiment. 
We see evidence of this process and measure the branching ratio 
as ${\cal B}(B_s^0\rightarrow D_s^{(*)}D_s^{(*)})= 0.035\pm0.010({\rm stat})\pm0.011({\rm syst})$. 
Based on this measurement and under certain theoretical assumptions, 
mixing and CP violation information in the $B_s^0$ meson system are extracted.
This is the first single measurement that demonstrates a non zero width difference in the 
$B_s^0$ system at greater than $3\sigma$ significance.
In particular, in the absence of NP, the fractional width difference is derived 
as $\Delta\Gamma_s^{\rm CP}/\Gamma_s = 0.072\pm0.021({\rm stat})\pm0.022({\rm syst})$.

%
We thank the staffs at Fermilab and collaborating institutions, 
and acknowledge support from the 
DOE and NSF (USA);
CEA and CNRS/IN2P3 (France);
FASI, Rosatom and RFBR (Russia);
CNPq, FAPERJ, FAPESP and FUNDUNESP (Brazil);
DAE and DST (India);
Colciencias (Colombia);
CONACyT (Mexico);
KRF and KOSEF (Korea);
CONICET and UBACyT (Argentina);
FOM (The Netherlands);
STFC (United Kingdom);
MSMT and GACR (Czech Republic);
CRC Program, CFI, NSERC and WestGrid Project (Canada);
BMBF and DFG (Germany);
SFI (Ireland);
The Swedish Research Council (Sweden);
CAS and CNSF (China);
and the
Alexander von Humboldt Foundation (Germany).
%

\end{document}